\include{Commands}
\documentclass[11pt]{article}
\usepackage{arxiv}
\usepackage{geometry}                
\usepackage{graphicx,bm}
\usepackage{epstopdf, epsfig}
\usepackage{xcolor,color}
\usepackage{amsmath}
\usepackage{mathtools}
\usepackage{amssymb}
\usepackage{amsthm}

\newcounter{defcounter}
\newenvironment{myequation}{%
\addtocounter{equation}{-1}
\refstepcounter{defcounter}

\begin{equation}}{\end{equation}}
\usepackage{hyperref}
\usepackage{cleveref}
\usepackage{comment}
\usepackage[color=green!20]{todonotes}

\usepackage[backend=bibtex,firstinits=true,style=numeric-comp,date=year,sorting=none]{biblatex}
\renewbibmacro{in:}{}
\AtEveryBibitem{\clearlist{language}}
\title{Formation of protein-mediated tubes is governed by a snapthrough transition }
\author{Arijit Mahapatra
and Padmini Rangamani$^*$ \\
Department of Mechanical and Aerospace Engineering, University of California San Diego,\\
9500 Gilman Drive, La Jolla, CA 92093, U.S.A.\\
Email addresses of corresponding authors: 
\href{prangamani@ucsd.edu}{\color{blue}{prangamani@ucsd.edu}}
}

\begin{document}

\maketitle

\begin{abstract}
   Plasma membrane tubes are ubiquitous in cellular membranes and in the membranes of intracellular organelles.
   They play crucial roles in trafficking, ion transport, and cellular motility. 
   The formation of plasma membrane tubes can be due to localized forces acting on the membrane or by curvature-induced by membrane-bound proteins. 
   Here, we present a mathematical framework to model cylindrical tubular protrusions formed by proteins that induce anisotropic spontaneous curvature. 
   Our analysis revealed that the tube radius depends on an effective tension that includes contributions from the bare membrane tension and the protein-induced curvature.
   We also found that the length of the tube undergoes an abrupt transition from a short, dome-shaped membrane to a long cylinder and this transition is characteristic of a snapthrough instability.
   Finally, we show that the snapthrough instability depends on the different parameters including coat area, bending modulus, and extent of protein-induced curvature.
   Our findings have implications for tube formation due to BAR-domain proteins in processes such as endocytosis, t-tubule formation in myocytes, and cristae formation in mitochondria. 
  
\end{abstract}


\section{Significance Statement}
Cylindrical tubes are ubiquitous on cellular membranes, and many studies have focused on how forces can give rise to tubes. But, how do curvature-inducing proteins, in the absence of forces, form cylindrical tubes on lipid bilayers? The answer relies on minimizing the bending energy of the membrane to accommodate two different principal curvatures such that a cylindrical shape minimizes energy. Here, we develop such a model and show that the formation of an elongated cylindrical tube is accompanied by a snapthrough transition. Furthermore, we identify how this transition depends on different membrane properties. These findings have implications for our understanding of how tubes forms in different cellular processes, including trafficking, mitochondrial inner membrane cristae formation, and T-tubule formation in myocytes.

\section{Introduction}

The generation of membrane curvature is an ubiquitous phenomenon in cells, both at the plasma membrane and at organelle membranes \cite{mahapatra2021mechanics}.
These curved structures can be broadly classified into buds, pearled structures, and tubes \cite{Yuan2021-su,Alimohamadi18}.
Specifically, the formation of tubes on membranes is attributed to either the interaction of cellular membranes with proteins that induce curvature or to forces exerted on the membrane by cytoskeletal assembly and motor proteins among other mechanisms.
Force-mediated membrane tubulation is quite well-studied and the relationship between the applied force and membrane properties is well understood \cite{derenyi2002formation,roux2002minimal}.
However, there are many situations in cells, where cylindrical tubes are formed due to curvature-inducing proteins alone without any apparent force mechanisms \cite{hong2014cardiac,rabl2009formation}. 
In this study, we focus on such protein-mediated tube formation. 

\begin{figure}[htbp]
    \centering
    \includegraphics[width=0.99\textwidth]{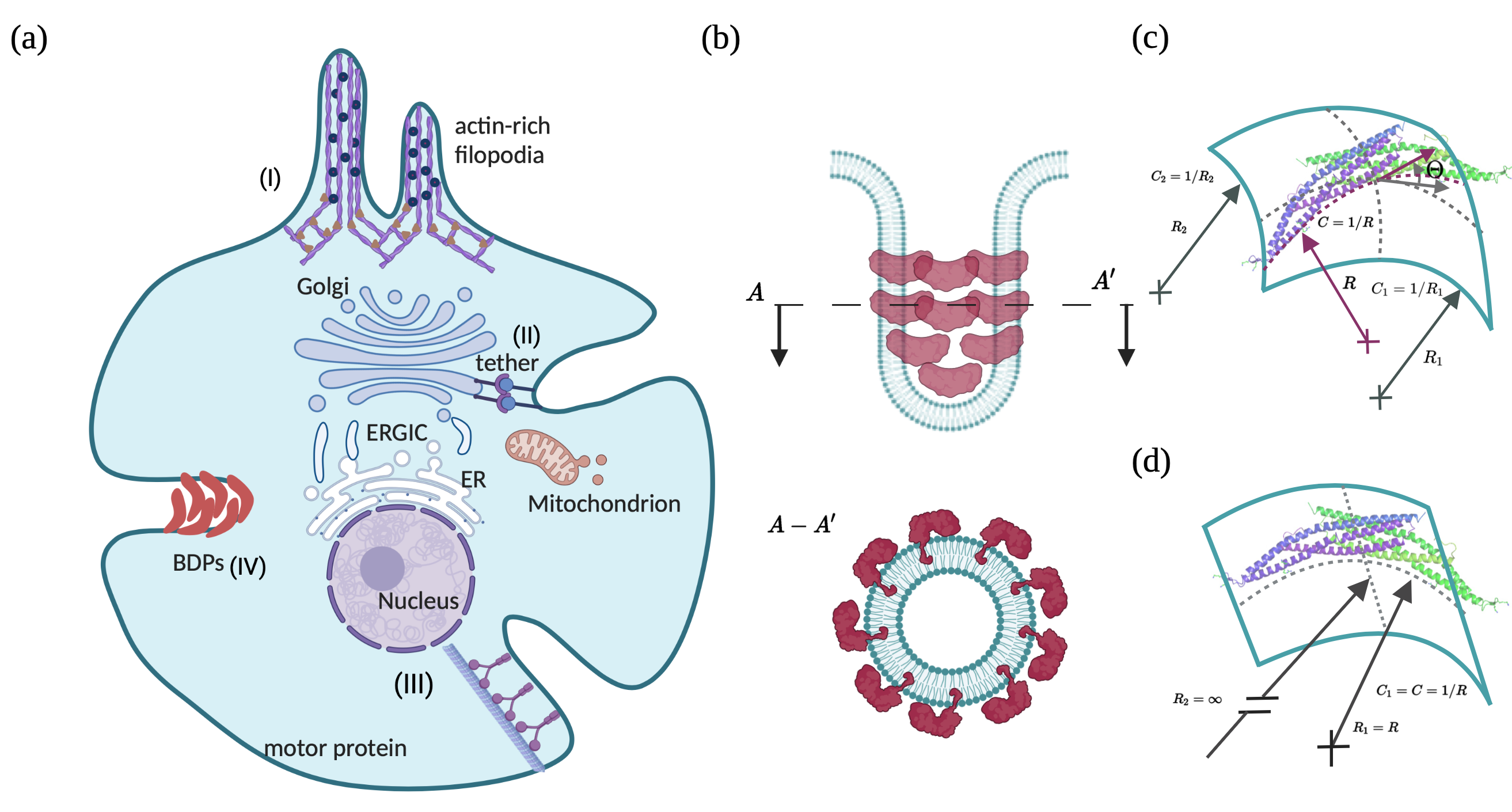}
    \caption{Schematic of protein-induced tube formation on cellular membranes. a) Mechanisms of tube formation in cellular membranes, (I) Tubes can be formed due to forces exerted on the membrane by the cytoskeleton, e.g. filopodial protrusion by actin filaments, (II) localized forces such as tethers and motors (III) can also lead to the formation of tubes. Tubules can also be formed by curvature-inducing proteins, such as BAR-domain proteins (BDPs) (IV). BDPs are know to form cylindrical tubules by inducing anisotropic curvature on the lipid bilayer. (b) Schematic depicting proteins that have anisotropic curvatures assembled on the cylindrical portion of a membrane tube.  The top subplot presents a side-view of the tube, and the bottom subplot shows the top view of the tube with BBPs. (c) The structure of a typical BAR-domain protein dimer \cite{PDB} induces two different curvatures in the parallel and perpendicular direction to enable a cylindrical deformation. (d) The particular case of BAR-domain dimer orientation causes curvature generation in one principal direction leading to a tubular protrusion. } \label{fig:schematic}
\end{figure}

Protein-mediated  tubes are essential components in numerous biophysical processes in a cell, \Cref{fig:schematic}a.
For example, in yeast, BAR(Bin/Amphiphysin/Rvs)-domain proteins play an important role in the elongated endocytic invaginations \cite{simunovic2015physics} and these proteins are known to dimerize and form curved structures. 
BAR-domain proteins such as BIN1 play an important role in the formation and maintenance of t-tubules in myocytes; these tubular structures are critical for excitation-contraction coupling \cite{hong2014cardiac,Hong15}.
Protein-decorated tubes are also important in intracellular organelles including the ER-Golgi complex \cite{Mollenhauer1998,raote2020tango1} and cristae in a mitochondria \cite{rabl2009formation}.
Experiments using BAR-domain proteins can generate tubular protrusions form from liposomes \textit{in vitro} \cite{Farsad2001,Takei1995,Takei1998,Takei1999}.
Stachowiak \textit{et al.} \cite{Stachowiak10} showed that protein crowding on giant unilamellar vesicles can cause tubulation, when a high density of proteins are attached.


From a theoretical standpoint, the mechanism of tube formation has been studied quite extensively using continuum mechanics and molecular dynamics simulations.
In the continuum models, the Helfrich energy \cite{Helfrich73} is used to capture the bending energy of the membrane. 
It is important to note that the formation of cylindrical tubes requires the accommodation of anisotropic curvatures.
The spontaneous curvature model accommodates spheres and pearls as energy-minimizing shapes because the minimum energy state of these geometries is such that both principal curvatures have the same value \cite{naito1995new}.
To obtain a cylinder, we need a formulation that accounts for one principal curvature to capture the cylinder's radius and the other principal curvature to equal zero (\Cref{fig:schematic}b). 
This can be done by recasting the Helfrich energy in terms of the mean curvature and the curvature deviator \cite{Iglic05,Iglic2007}. 
Such models have been proposed and simulations have been used to explore tubulation \cite{Iglic2013,Walani2014} and the impact of BAR-domain proteins on endocytosis \cite{Walani2015}.

While there are many models that address the issue of tubule formation, we found that the formation of a complete cylindrical tube --  defined here as a cylindrical membrane with a hemispherical cap -- is not a trivial mathematical or computational problem.
Specifically, we note that the use of the classical Helfrich energy with a spontaneous isotropic curvature will not admit solutions that will give rise to cylindrical tubes with a hemispherical cap. 
Furthermore, we identified that a comprehensive analysis energy landscape of tube formation in response to different mechanical properties of the membrane is currently lacking in the literature. 
In this work, we conducted a detailed mechanical analysis of tube formation due to membrane-bound proteins that induce anisotropic curvature using the curvature deviator version of the Helfrich model.
We found that the formation of membrane tubules due to anisotropic curvature is governed by a snapthrough instability that depends on the membrane tension and the bending modulus. 
We also establish scaling relationships between membrane tension, extent of anisotropic curvature, and the bending modulus and the tube dimensions.
Finally, we predict how the snapthrough instability and the bending energy landscape varies with different membrane parameters.

\section{Model development}

\begin{table}
\caption{Notation and list of symbols}
\label{table:Notation}
\centering
\begin{tabular}{lll}
\hline Notation & Description & Unit \\
\hline
   $W$ & Free energy density of the membrane  &  $\mathrm{pN/nm}$ \\
  $\kappa$ & Bending modulus of the membrane  & $\mathrm{pN \cdot nm}$ \\
    $\kappa_d$ & Deviatoric modulus of the membrane  & $\mathrm{pN \cdot nm}$ \\
    $\boldsymbol{n}$ & Surface normal   & 1 \\
$\boldsymbol{\zeta}$ & Orientation vector of proteins  & 1  \\
$\boldsymbol{\mu}$ & Binormal vector of protein $\boldsymbol{\mu}=\boldsymbol{n} \times \boldsymbol{\zeta}$ (see  \Cref{fig:Prb1}) & 1  \\
$a_{\alpha \beta },a^{\alpha \beta }$ & Metric tensor and its contravariant & -- \\
$b_{\alpha \beta },b^{\alpha \beta }$ & Curvature tensor and its contravariant & -- \\
  $ s$ & Arclength along the membrane & $\mathrm{nm}$ \\
  $ \psi$ & Angle made by surface tangent with the horizontal direction  & 1 \\
  $C$ & Curvature of the BAR-domain proteins & $\mathrm{nm}^{-1}$ \\

  $C_1,C_2$ & Protein-induced principal curvatures & $\mathrm{nm}^{-1}$ \\
$c_1,c_2$ & Principal curvatures of the membrane & $\mathrm{nm}^{-1}$ \\
$H$ & Mean curvature & $\mathrm{nm}^{-1}$ \\
$D$ & Deviatoric curvature & $\mathrm{nm}^{-1}$ \\
$\boldsymbol{T}$ & Surface traction & $\mathrm{pN/nm}$ \\
  $p$ & Normal pressure acting on the membrane & $\mathrm{pN/nm^2}$ \\
$\lambda$ & Membrane tension & $\mathrm{pN/ nm}$ \\
   &   & \\
\hline
\end{tabular}
\end{table}

\begin{table}
\caption{List of parameters}
\label{table:parameters}
\centering
\begin{tabular}{lll}
\hline Notation & Description & Range \\
\hline
$C_0$ & Spontaneous mean curvature &  $0.01 \textendash 0.02~\mathrm{nm}^{-1}$ \cite{yin2009simulations}  \\
$D_0$ & Spontaneous deviatoric curvature & $0.01 \textendash 0.02~\mathrm{nm}^{-1}$ \cite{yin2009simulations}   \\
 $\kappa(=\kappa_d)$ & Bending modulus of the membrane  & $21\textendash336~\mathrm{pN\cdot nm}$ \cite{schneider1984thermal,fowler2016membrane} \\
$\lambda_0$ & Membrane tension at the boundary & $0.001 \textendash 0.1 ~\mathrm{pN/nm}$ \cite{Lipowsky12,rangamani2022many} \\
$a_{\mathrm{coat}}$ & Coat area of the protein & $2.5 \times 10^{4} \textendash 5 \times 10^{4} ~\mathrm{nm}^{2}$ \cite{Hassinger17} \\
 &   & \\
\hline
\end{tabular}
\end{table}

From a geometric perspective, a membrane can be described from the mean ($H$) and deviatoric ($D$) curvature, as given by 
\begin{equation}
    H= \frac{c_1+c_2}{2}, D=\frac{c_1-c_2}{2},
\end{equation}
where $c_1$ and $c_2$ are the principal curvatures of the surface.
Specific combinations of these curvatures define characteristic surfaces such as spherical buds, tubes, and catenoids.
BAR-domain proteins (BDPs) are a classic example of curvature-inducing proteins that result in membrane tubes. This family of proteins can be treated as one-dimensional curved proteins \cite{PDB} that induce anisotropic curvature to the membrane (\Cref{fig:schematic}c,d, also see (\Cref{fig:Prb1})) \cite{Iglic05,Iglic05a,Walani2014}.
If $C_1$ and $C_2$ are the curvatures induced by the proteins in the two principal directions,
the spontaneous mean ($C_0$) and deviatoric ($D_0$)  are given by
\begin{equation}
    C_0= \frac{C_1+C_2}{2}, D_0=\frac{C_1-C_2}{2}.
\end{equation}

Note that the principal components of the induced curvatures, $C_1$ and $C_2$, will depend on the value of curvature $C$ of the BAR-domain proteins and the orientation angle $\Theta$ (\Cref{fig:schematic}d)). 
For such orientation, the spontaneous mean ($C_0$) and deviatoric curvature ($D_0$) relate to the curvature of the proteins ($C$) and their orientation angle $\Theta$ with the following relationship \cite{Kabaso2011attachment,Iglic2013,Perutkova2010}
\begin{equation}
    C=\left(\frac{C_1+C_2}{2} \right)+\left(\frac{C_1-C_2}{2} \right)\cos(2\Theta)=C_0+D_0\cos (2\Theta).
\end{equation}
In the example shown in \Cref{fig:schematic}d, BDPs orient circumstantially along a cylindrical membrane ($C_2=0$) with orientation angle $\Theta=0$.
In this case, spontaneous mean and deviatoric curvatures is given by 
\begin{equation}
    C_0=\frac{C}{2} \quad \text{and} \quad D_0=\frac{C}{2}.
\end{equation}
\begin{figure}[htbp]
    \centering
    \includegraphics[width=0.8\textwidth]{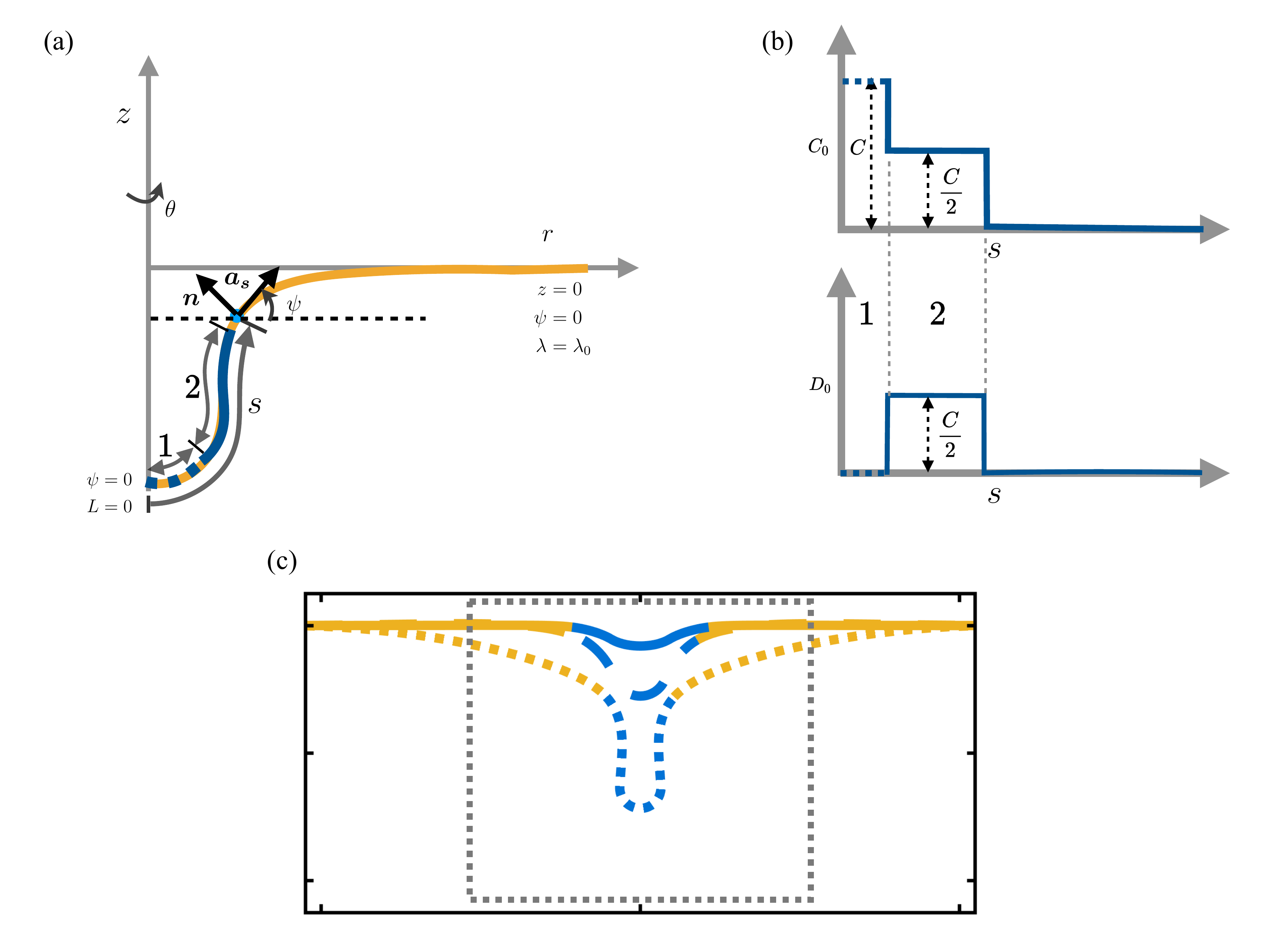}
    \caption{Membrane representation in axisymmetry with protein distribution area and boundary conditions. (a) Computational domain showing the axisymmetric coordinates, location of the protein coat, and boundary conditions. (b) Input distribution of spontaneous mean and deviatoric curvatures on the membrane along its arclength. (c) Representative deformed membrane configuration; the dashed lines show the region in the $r-z$ plane we show in the subsequent figures.} \label{fig:figure2}
\end{figure}
\subsection{Assumptions}
We assumed that membrane thickness and change in thickness are negligible compared to bending, and thus the Helfrich description is valid in this case.
The membrane is elastic in bending, and the bending rigidity is assumed to be uniform across the solution domain (\Cref{fig:figure2}a).
We also assumed that the membrane is incompressible and in mechanical equilibrium. 
We do not consider the molecular details of the protein-induced curvature but rather use spatial distributions of $C_0$ and $D_0$ as inputs to the model. 
To capture the hemispherical cap and the cylindrial tube regions of the membrane, we assumed that the spontaneous mean and deviatoric curvature induced by membrane-bound BDPs are divided into two regions (see \Cref{fig:figure2}),\\
\textbf{i.} at the tip of the coat, a spontaneous mean curvature cap $C_0=C$ with zero spontaneous deviatoric curvature ($D_0=0$) (region 1 in \Cref{fig:figure2}a,b),\\
\textbf{ii.} the rest of the coat with equal values of spontaneous mean and deviatoric curvature $C_0=D_0=C/2$ (region 2 in \Cref{fig:figure2}a,b).\\
Note that this type of spontaneous curvature distribution arises when BDPs are  aligned axially in the hemispherical region (region 1 in \Cref{fig:figure2}a) and circumstantially in the cylindrical region (region 2 in \Cref{fig:figure2}a).
This orientation arises from the minimization of bending energy of the BAR proteins \cite{Perutkova2010,tozzi2021theory}. 
However, the spontaneous curvature at the tip of the coat might be induced by other curvature-generating proteins but we do not consider the molecular details here. 

\subsection{Free Energy of the membrane}
The free energy of the membrane is similar to the elastic energy given by the Helfrich Hamiltonian \cite{Helfrich73} modified to include the deviatoric components \cite{Kabaso2011attachment,Iglic2013,Walani2014}
\begin{equation}
\label{eqn:energy}
    W=\kappa (H-C_0)^2+\kappa_d (D-D_0)^2,
\end{equation}
where $\kappa,\kappa_d$ are the bending rigidities, $H$ and $D$ are the mean and deviatoric curvatures of the membrane, and $C_0, D_0$ are the spontaneous isotropic and deviatoric curvatures induced by proteins respectively.

\subsection{Governing equations}
Local stress balance on the membrane gives us \cite{Steigmann99,Steigmann03,Rangamani13} 
\begin{equation}
 \boldsymbol{T}^{\alpha}_{;\alpha}+p\boldsymbol{n}=\boldsymbol{0}\,,
\label{eqn:eqm1}
\end{equation}%
where $p$ is normal pressure on the membrane and $\boldsymbol{T}$ is the surface stress tensor. \Cref{eqn:eqm1} results in force balance equations in the tangential and normal direction of the membrane (see \Cref{sec:si_sec1_2} in the supporting information).
The tangential force balance relation reduces to \cite{Steigmann03,Rangamani13,Hassinger17,Walani2014} 
\begin{equation}
\label{eqn:tang_force_bal}
\lambda_{, \alpha}=-W_{,\alpha|exp}, 
\end{equation}
where $\lambda$ is the Lagrange multiplier for area constraints.
Physically it represents the tension in the plane of the membrane \cite{Rangamani14,mahapatra2020transport,mahapatra2021curvature,rangamani2022many}. 
$W_{,\alpha|exp}$ in \Cref{eqn:tang_force_bal} represents the explicit dependence of the energy on the coordinates
\cite{Steigmann03,Rangamani13}.
The normal force balance relation depicts the equilibrium shape of the membrane and is given by \cite{Walani2014}
\begin{equation}
\begin{split}
\frac{1}{2}  [W_D(\zeta^{\alpha}\zeta^{\beta}-\mu^{\alpha}\mu^{\beta})]_{;\beta \alpha} + \frac{1}{2} W_D (\zeta^{\alpha}\zeta^{\beta}-\mu^{\alpha}\mu^{\beta})b_{\alpha \gamma} b^{\gamma}_{\beta}+
\Delta (\frac{1}{2}W_H)+ \\ (W_K)_{;\beta \alpha}(2H {a}^{\beta \alpha}- {b}^{\beta \alpha})+ W_H(2H^2-K)+2H(KW_K-W)-2H\lambda=p.
\end{split}
\end{equation}
Here $\boldsymbol{\zeta}$ is the direction of orientation of the BDP's, $\boldsymbol{\mu}=\boldsymbol{n}\times \boldsymbol{\zeta}$, with $\boldsymbol{n}$ as the unit surface normal \cite{Walani2014}.

\subsection{Governing equations in axisymmetry}
In polar coordinates, the geometry of the membrane can be parameterized by 
\begin{equation}
    \boldsymbol{r}(r,z,\theta)=\boldsymbol{r}(s,\theta),
\end{equation}
where $s$ is the arclength, and $\theta$ angle in the azimuthal direction. 
In the limit of axisymmetry, $\frac{\partial ( \cdot )}{\partial \theta}=0$ and the membrane can be parameterized by the arclength $s$ only, as shown in \Cref{fig:figure2}. 

In axisymmetry, the tangential force balance relation in  \Cref{eqn:tang_force_bal} reads as
\begin{equation}
\label{eqn:tang_force_bal_axi}
\lambda^{\prime}=2 \kappa (H-C_0)C_0^{\prime}+2 \kappa_d (D-D_0)D_0^{\prime}, 
\end{equation}
where $(\cdot)^{\prime}=\frac{\partial (\cdot)}{\partial s}$. 
The normal force balance in \Cref{eqn:tang_force_bal} relation simplifies to
\begin{equation}
\begin{aligned}
p=& \frac{L^{\prime}}{r}+2 \kappa (H-C_0)\left(2 H^{2}-K\right)-2 H\left(W+\lambda-2 \kappa_d D(D-D_0)\right)
\end{aligned}
\end{equation}
where,
\begin{equation}
\label{eqn:L}
\frac{L}{r}=\left[\kappa (H-C_0)-\kappa_d (D-D_0)\right]^{\prime} - 2 \kappa_d (D-D_0)\frac{\cos \psi}{r}, 
\end{equation}
is the normal component of traction on the membrane.
We note that all of these equations are presented in \cite{Walani2014} but there is a missing term in $L$ in that work; a complete derivation is presented in \Cref{sec:si_gov_eqn} of the supporting information.  

\subsection{Numerical methods}
We solved the system of equations (\Cref{myeqn:tang_force_bal_axi_nd} to \Cref{myeqn:L}) numerically to get the equilibrium shape of the membrane for a coat of protein at the center of an axisymmetric area patch. 
The solution domain is presented in \Cref{fig:figure2}a, along with the input protein coat and the boundary conditions (\Cref{fig:figure2}a,b).
The protein coat includes both the spontaneous mean curvature cap and a combination of mean and deviatoric spontaneous curvature in the rest of the coat region (\Cref{fig:figure2}).
Note that we introduced a shape variable $\psi$ which denotes the angle made by the tangent from its radial plane. 
The membrane is clamped at the domain boundary, where both the displacement and the angle $\psi=0$. The membrane tension is also prescribed at the boundary. 
At the pole, $\psi$ is taken to be zero, which indicates the smoothness at the center of the membrane. 
$L$ is set to zero, indicating that there is no pulling force acting at the center.
This distinguishes our work from force-mediated tubulation that was studied in other works \cite{derenyi2002formation,roux2002minimal}. 
To solve the system of equations, we used MATLAB-based bvp4c, a finite difference-based ODE solver with fourth-order accuracy (MATLAB codes are available \href{https://github.com/Rangamani-Lab/arijit_deviatoric_tube_2022}{ https://github.com/Rangamani-Lab/arijit$\_$deviatoric$\_$tube$\_$2022}).
We used a nonuniform grid ranging from 1000 to 10000 points, with the finer grid towards the center.
We used a large domain size of $10^6~\mathrm{nm}^2$ to avoid boundary effects but we show the results focusing on the membrane deformation (region enclosed by the dashed line in \Cref{fig:figure2}c).


\section{Results}
Here we establish the effect of different parameters on the formation of a cylindrical membrane tube and show that membrane tension and bending rigidity govern the energy landscape of tubule formation due to proteins inducing deviatoric curvature. 
Using numerical simulations, we show that this energy landscape is governed by a snapthrough instability, and different parameters such as area of induced curvature and extent of induced curvature tune the location of the snapthrough transition. 

\subsection{Membrane tension and bending rigidity regulate the tubule characteristics in the presence of curvature inducing proteins.}
\begin{figure}[htbp]
    \centering
    \includegraphics[width=0.99\textwidth]{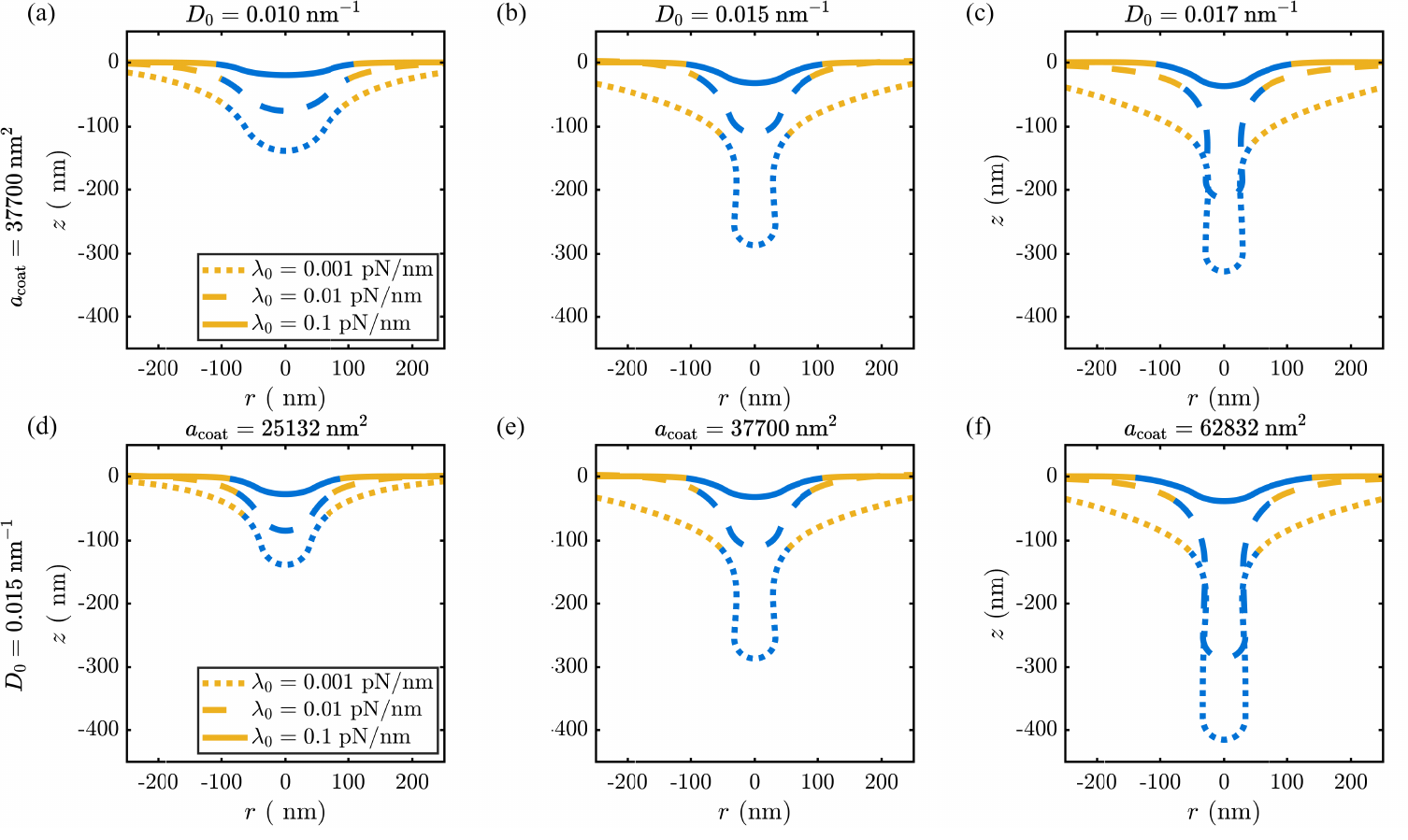}
    \caption{Membrane tension governs tubulation. (a-c) Equilibrium shapes of the membrane for 3 different spontaneous deviatoric curvatures, (a) $D_0=0.010 \mathrm{~nm^{-1}}$, (b) $D_0=0.015 \mathrm{~nm^{-1}}$, and (c) $D_0=0.017 \mathrm{~nm^{-1}}$.
    For each value of deviatoric curvature, we investigated tubule formation at three different values of tension (0.001, 0.01, 0.1 $\mathrm{~pN/nm}$). 
    The area of the protein coat was fixed at $37700 ~ \mathrm{~nm}^2$ and bending rigidity of the membrane was taken to be $42 \mathrm{~pN\cdot nm}$, which is roughly $10 \mathrm{~k_BT}$. (d-f) Equilibrium shape of the membrane for 3 different protein coats areas, (d) $a_{\text{coat}}=25312 \mathrm{~nm^2}$,  (e) $a_{\text{coat}}=37700 \mathrm{~nm^2}$, and (f) $a_{\text{coat}}=62832 \mathrm{~nm^2}$ for three different values of tension (0.001, 0.01, 0.1 $\mathrm{pN/nm}$). The spontaneous deviatoric curvature was fixed at $0.015~ \mathrm{nm}^{-1}$ and bending rigidity of the membrane $42 \mathrm{~pN\cdot nm}$, which is roughly $10 \mathrm{~k_BT}$.} \label{fig:tension_variation}
\end{figure}

We previously showed that membrane tension regulates the formation of spherical vesicles \cite{Hassinger17} and it is well-established that membrane tension governs the force-length relationships of force-mediated tube formation \cite{derenyi2002formation}. 
How does tension regulate the formation of protein-induced membrane tubules?
To answer this question, we first fixed the coat area and the deviatoric curvature and varied membrane tension (\Cref{fig:tension_variation}a-c). 
We found that, for a fixed coat area of $37,700 \mathrm{~nm^2}$, low values of deviatoric curvature $D_0=0.010 \mathrm{~nm^{-1}}$ were not sufficient to induce the formation of cylindrical structures for the range of membrane tension tested ($\lambda_0=0.001, 0.01, 0.1 \mathrm{~pN/nm}$). 
The deformations in these cases were dome-shaped without any location with multiple $z$ values for a given $r$.
However, when the value of the protein-induced curvature was increased to $D_0=0.015 \mathrm{~nm^{-1}}$ (\Cref{fig:tension_variation}b) and to $D_0=0.017 \mathrm{~nm^{-1}}$ (\Cref{fig:tension_variation}c), we found that lower values of tension promoted the formation of cylindrical tubules.
These results indicate that for a fixed coat area, the combination of induced-curvature and tension regulate the transition from dome-shaped deformations to cylindrical deformations. 
We then investigated how the area of the coat would affect these transitions (\Cref{fig:tension_variation}d-f) for a fixed amount of induced curvature, $D_0=0.015 \mathrm{~nm^{-1}}$.
We found that lower coat area favored dome-shaped deformations even at low tension (\Cref{fig:tension_variation}d) but higher coat area promoted the formation of cylindrical tubes (\Cref{fig:tension_variation}e, f) and the length of the tube increased with decreasing tension. 

\begin{figure}[htbp]
    \centering
    \includegraphics[width=0.99\textwidth]{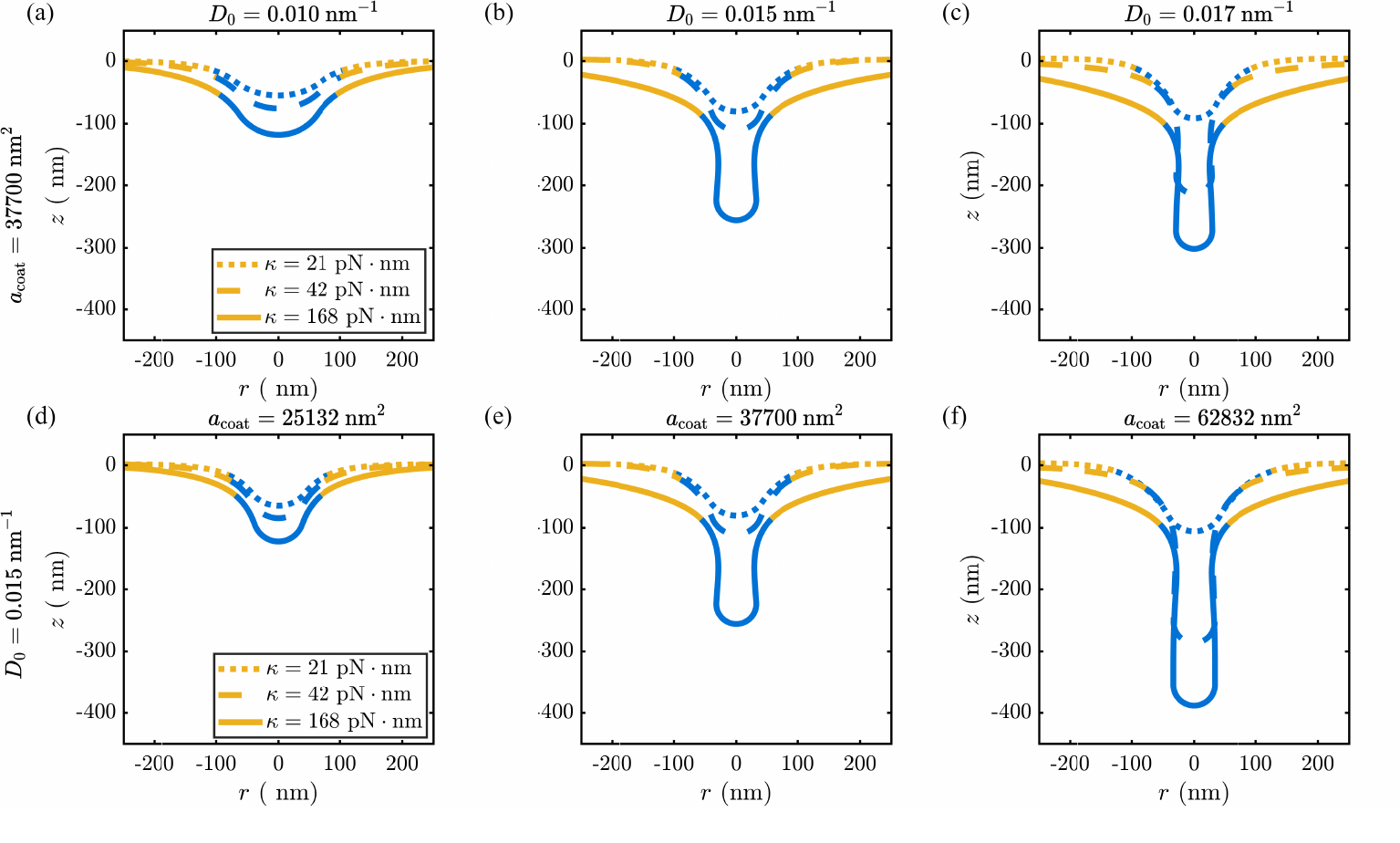}
    \caption{Bending rigidity impacts membrane tubulation. (a-c) Equilibrium shape of the membrane for 3 different spontaneous deviatoric curvatures, (a) $D_0=0.010 \mathrm{~nm^{-1}}$, (b) $D_0=0.015 \mathrm{~nm^{-1}}$, and (c) $D_0=0.017 \mathrm{~nm^{-1}}$.
    For each value of deviatoric curvature, we investigated tubule formation at three different values of bending ridigity (21, 42, and 168 $\mathrm{pN\cdot nm}$). The area of the protein coat was fixed at $37700 ~ \mathrm{nm^2}$ and the tension was fixed at 0.01 $\mathrm{~pN/nm}$.
    compared with three different values of bending rigidities (21, 42, 168 $\mathrm{pN \cdot nm}$). 
    (d-f) Equilibrium shape of the membrane for 3 different protein coats areas, (d) $a_{\text{coat}}=25312 \mathrm{~nm^2}$,  (e) $a_{\text{coat}}=37700 \mathrm{~nm^2}$, and (f) $a_{\text{coat}}=62832 \mathrm{~nm^2}$ for three different values of bending ridigity (21, 42, and 168 $\mathrm{pN\cdot nm}$). The area of the protein coat was fixed at $37700 ~ \mathrm{nm^2}$ and the tension was fixed at 0.01 $\mathrm{pN/nm}$.} \label{fig:bending_variation}
\end{figure}

We next asked if the bending rigidity of the membrane also played a role in the formation of cylindrical tubes. 
The characteristic length scale of the membrane is $\sqrt{\kappa/\lambda}$ and therefore, if tension were to be fixed constant but bending rigidity were to change, we would expect to see similar results. 
We fixed the coat area and protein-induced curvature and varied the bending ridigity of the membrane for a fixed value of membrane tension (\Cref{fig:bending_variation}).
For low values of protein-induced curvature, $D_0=0.010 \mathrm{~nm^{-1}}$, the membrane deformation remained dome-shaped for all values of membrane bending rigidity (\Cref{fig:bending_variation}a). 
Increasing the protein-induced curvature increased the deformation and resulted in cylindrical tubules for higher values of bending rigidity (\Cref{fig:bending_variation}b,c).
We observed the same trend for fixed protein-induced curvature and increasing coat area (\Cref{fig:bending_variation}d-f). 
Smaller coat areas maintained a dome-shaped deformation but larger coat areas resulted in cylindrical tubes as the bending rigidity increased.

It is not surprising that decreasing membrane tension increased tubulation and this is consistent with experimental observations \cite{barooji2016dynamics}. 
However, it may seem counterintituitive that higher membrane modulus promotes tubulation.
This result can be understood as follows -- in all our simulations, we already prescribe the spontaneous curvature \textit{a priori} and it does not evolve with the shape of the membrane. 
In other words, the membrane is conforming to a shape that minimizes the energy for a given value of $D_0$ and the prescribed boundary conditions.
For a tensionless membrane, with free boundaries, the membrane will conform to a cylinder such that it takes on the curvature that is prescribed. 
Since we have prescribed a boundary condition with a finite tension $\lambda_0$, this will promote flattening of the membrane and an increase in bending rigidity will oppose this flattening.
Thus, the competition between the bending energy trying to form a cylinder versus the edge tension trying to flatten the membrane results in increased bending rigidity promoting cylindrical tube formation.

\begin{figure}[htbp]
    \centering
    \includegraphics[width=0.95\textwidth]{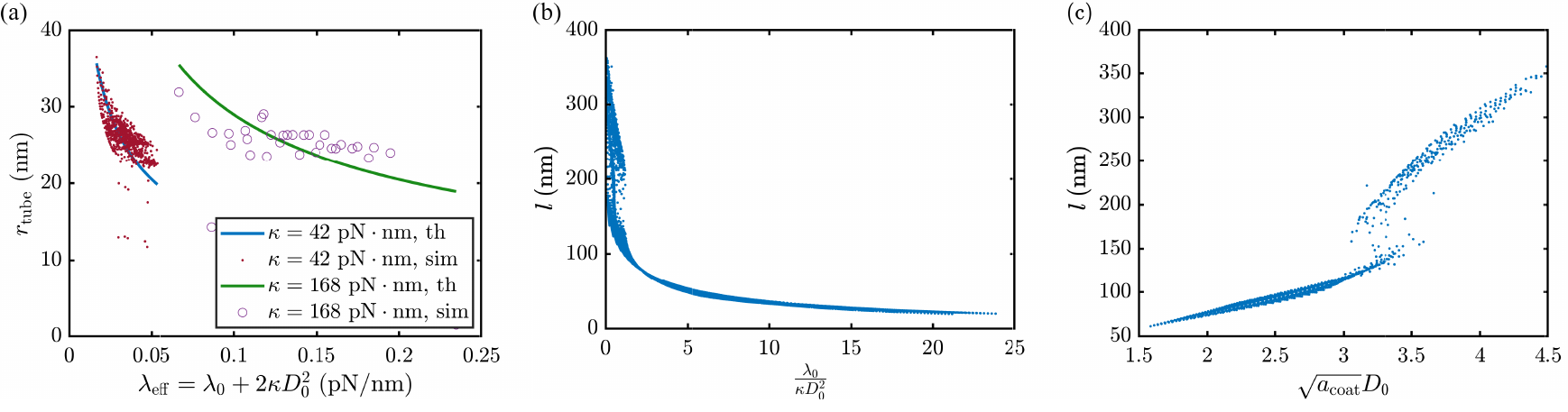}
    \caption{Scaling relationships for tube radius and tube length. (a) Effective tension ($\lambda_{\mathrm{eff}}=\lambda_0+2\kappa D_0^2$) governs the radius of the tubular protrusions. A comparison of tube radii from theoretical prediction and numerical simulations for two different values of $\kappa$ is shown. The solid curves are from the theoretical values in \Cref{eqn:theoretical} and the dots are from numerical simulations for different values of $\lambda_0$ and $D_0$. The simulations deviate from the theoretical predictions because of edge conditions and the hemispherical cap. 
    (b) Length of the tube scales with $\lambda_0/(\kappa D_0^2)$ and with $\sqrt{a_{\mathrm{coat}}}D_0$ (c). In both these cases, a transition is observed in the nature of the curve around $l=150-200 \mathrm{nm}$, suggesting a transition in the energy of the tube. For panel (b) protein coat area was fixed at 3.77$\times 10^4$ $\mathrm{nm}^2$. For panel (c), bending rigidity was fixed at  $\kappa = 42~\mathrm{pN \cdot nm}$ and tension at $\lambda_0$ of $0.01~ \mathrm{pN/nm}$.  }
    \label{fig:dimensionless}
    
\end{figure}

\begin{figure}[h] 
    \centering
    \includegraphics[width=\textwidth]{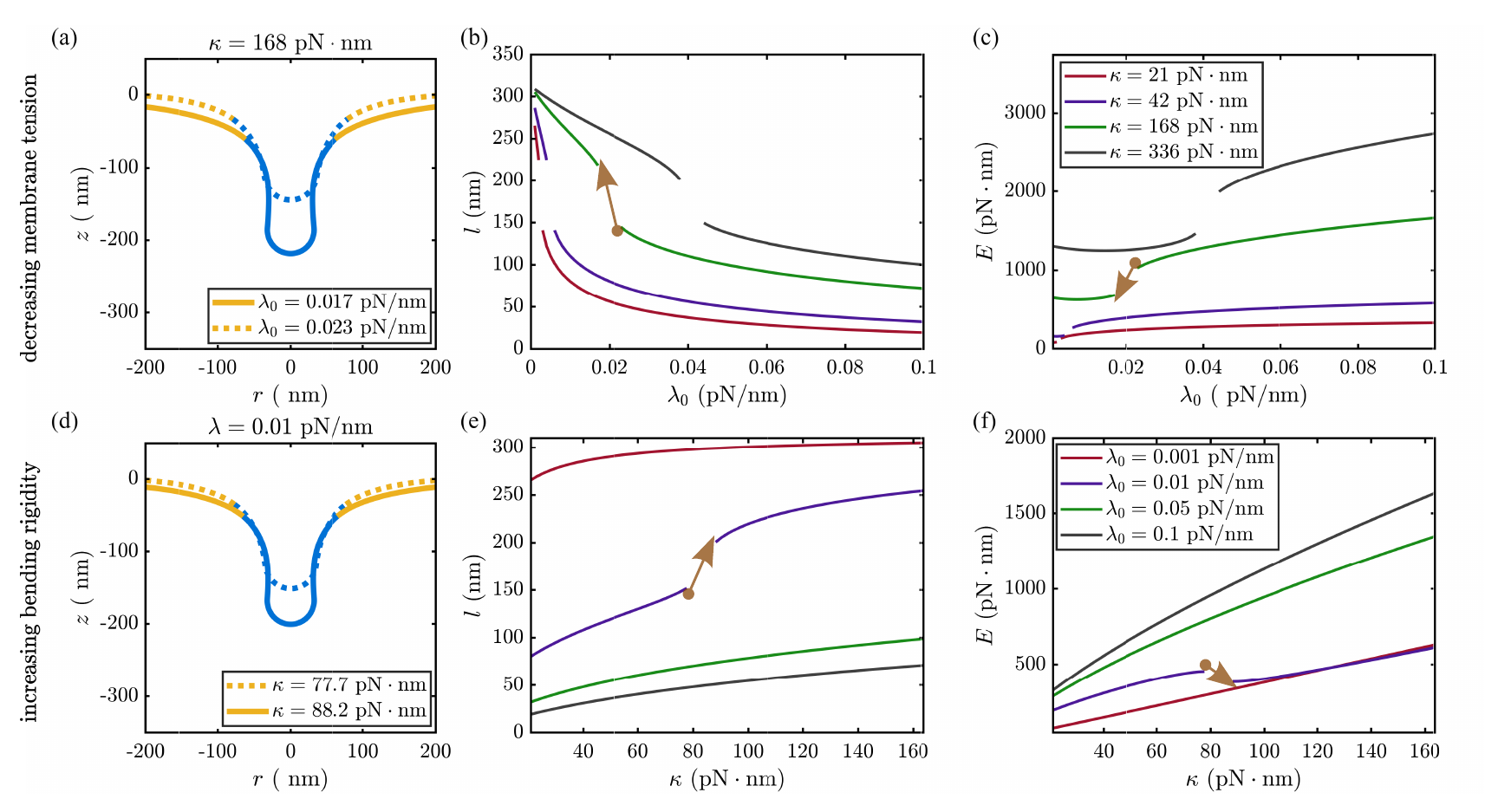}
    \caption{The dome-to-cylinder transition is governed by a snapthrough instability. 
    (a,b,c) Snapthrough instability caused by decreasing membrane tension membrane tension with $a_{\mathrm{coat}}=3.77\times10^4{~\mathrm{nm}^2}$ and $D_0=0.015 ~\mathrm{nm}^2$ with the landscape of energy and membrane deformation. (a) The shape of the membrane before (dashed) and after (solid line) after the transition is shown.
    Both the length of the tubular structure (b) and the bending energy (c) have discontinuities across the snapthrough for different values of bending rigidity $\kappa$.
    (d,e,f) Snapthrough instability caused by increasing bending rigidity $\kappa$ with $a_{\mathrm{coat}}=3.77\times10^4{~\mathrm{nm}^2}$ and $D_0=0.015 ~\mathrm{nm}^2$ with the landscape of energy and membrane deformation. the morphology of the membrane changed from dome to a cylinder shape (d). The length of the tubular structure (e) and the bending energy (f) are shown with bending rigidity for different values of membrane tension. Snapthrough occurs for $\lambda_0=0.01 \mathrm{~pN/nm}$. 
    } \label{fig:snapthrough}
\end{figure}

\subsection{Scaling relationships for tube radius and tube length}
Given that the formation of a cylindrical tubule depends on multiple parameters such as tension, bending modulus, and the extent of curvature induced by the proteins, we next derived scaling relationships to relate these different mechanical properties to the geometry of the cylinder. 
Motivated by the analysis in \cite{derenyi2002formation}, we considered the free energy of a cylindrical membrane of radius $R$ and length $l$ whose entire area is covered by proteins that induce a curvature of $D_0$. 
We ignore the hemispherical cap and the edge conditions. 
In this case, the free energy of a cylinder is given by 
\begin{equation}
E= 4 \pi R l \bigg\{ \kappa \left(\frac{1}{2R}- D_0 \right)^2   \bigg\} + \lambda_0 2\pi R l.
\end{equation}
We can obtain radius by minimizing the energy with the radius of the tube $R$. 
The variation of energy $E$ with $R$ is given by
\begin{equation}
\begin{split}
\frac{\partial E}{ \partial R} & =   4 \pi R l \bigg\{ 4 \kappa \left(\frac{1}{2R}- D_0 \right) \left(-\frac{1}{2R^2} \right)  \bigg\} + \lambda_0 2\pi  l \\
&= 4 \pi \kappa l \left( \frac{1}{2R} -D_0 \right) \left( \frac{1}{2R} -D_0 -\frac{1}{R} \right)+\lambda_0 2 \pi l \\
&= -  4 \pi \kappa l  \left( \frac{1}{4R^2} -D_0^2 \right) + \lambda_0 2 \pi l
\end{split}
\end{equation}
Energy minimization results in
\begin{eqnarray}
R&=\frac{1}{\sqrt{\frac{2 \lambda_0}{\kappa}+4D_0^2}}\\
& = \sqrt{\frac{\kappa}{2\lambda_{eff}}}.
\label{eqn:theoretical}
\end{eqnarray}
where $\lambda_{eff}=\lambda_0+2D_0^2 \kappa$ is an effective tension. 
This relationship is reminiscent of the radius of a tube extruded using a pulling force, $R=\sqrt{\kappa/2\lambda}$ \cite{derenyi2002formation}, with an effective tension $\lambda_{eff}$ that accounts for the tension contributions from the protein coat \cite{Rangamani14,Lipowsky12,mahapatra2020transport,mahapatra2021curvature}. 
We compared our theoretical prediction against simulations for two different values of $\kappa$ (\Cref{fig:dimensionless}a). 
We found that the values from the simulations were very close to the theoretically predicted curves; the deviations are a result of the boundary conditions and the hemispherical cap in the simulation geometry.

\begin{figure}[h]
    \centering
    \includegraphics[width=\textwidth]{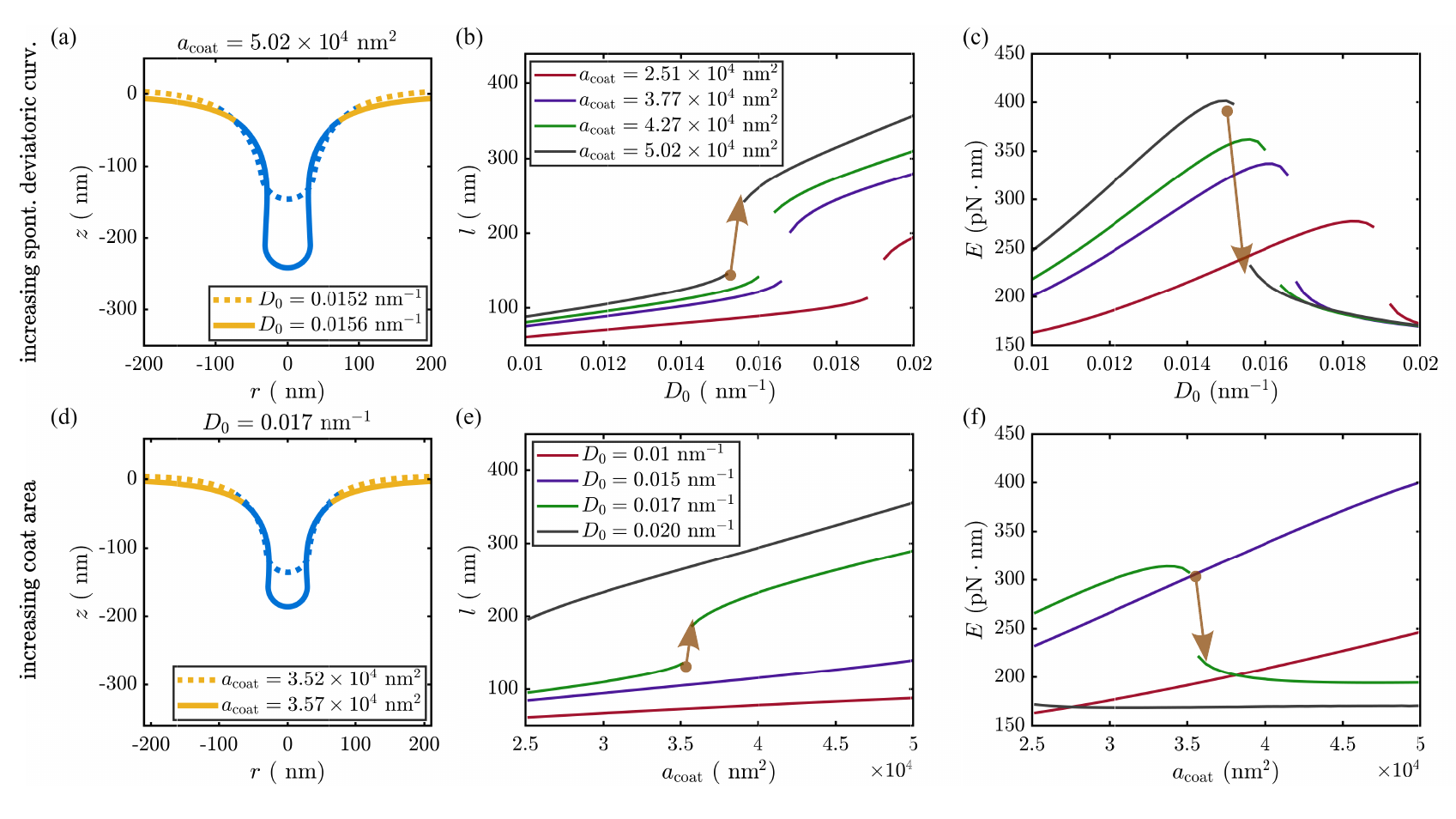}
    \caption{The role of spontaneous deviatoric curvature and the coat area on snapthrough instability. (a,b,c) Snapthrough instability caused by increasing spontaneous deviatoric curvature with $\kappa=\kappa_d=42 ~\mathrm{pN/nm} $ and $\lambda_0=0.01 ~\mathrm{pN/nm}^2$ with the landscape of energy and membrane deformation. Both the length of the tubular structure (b) and the bending energy (c) experienced discontinuities across the snapthrough, and are shown for 4 different values of bending rigidity $\kappa$. The morphology of the membrane changed from dome to cylinder shape (a). (d,e,f) Snapthrough instability caused by increasing coat area of protein with $\kappa=\kappa_d=42 ~\mathrm{pN/nm} $ and $\lambda_0=0.01 ~\mathrm{pN/nm}^2$ with the landscape of energy and membrane deformation. The length of the tubular structure (e) and the bending energy (f) are shown with bending rigidity for different values of membrane tension. For $\lambda_0=0.01 \mathrm{~pN/nm}$, a discontinuity is observed where the morphology of the membrane changed from dome to cylinder shape (d).} \label{fig:D0_a0}
\end{figure}

We note that there is no equivalent relationship for force on the tip of the tube because our energy does not contain a force term. Therefore, we cannot obtain a scaling relationship for the length of the tube with the forces applied at the tip of the tube.
Instead, we sought dimensionless numbers from the governing equations that could describe the length of the tube in a reduced parameter space. 
We constructed these dimensionless numbers with the help of the Buckingham-Pi theorem as function of characteristic scales.
We found that the length of the tube for different conditions collapses to a line as a function of  $\frac{\lambda_0}{\kappa D_0^2}$ (\Cref{fig:dimensionless}b). 
These simulations were performed for different values of $\lambda_0$, $\kappa$, and $D_0$ keeping coat area $a_{\mathrm{coat}}$ constant. 
We further observe a transition in the length around $200 \mathrm{~nm}$ indicating a change in nature of the system. 
We also observed that the length of the tube collapses onto a line with another  dimensionless number $\sqrt{a_{\mathrm{coat}}}D_0$ (\Cref{fig:dimensionless}c). 
This is because the tube length depends not only on the membrane parameters but also on the area of the coated region. 
Again, we observe a transition in the nature of the length of the tube around $150-200~ \mathrm{nm}$, depicted by a decreasing density of points.

\subsection{Formation of cylindrical tubes is governed by a snap-through instability}
In our simulations, we observed that the transition from a dome-shaped cap to a rigid, cylindrical tube appeared as an abrupt transition (\Cref{fig:tension_variation,fig:bending_variation}) and a transition accompanying the elongation of the tube (\Cref{fig:dimensionless}). 
We conducted further simulations around regions of these abrupt transitions and investigated the energy landscape (\Cref{fig:snapthrough}). 
As before, we fixed the bending rigidity and decreased the membrane tension (\Cref{fig:snapthrough}a-c). 
We found that the dome-to-cylinder transition took place between $\lambda_0=0.023 \mathrm{~pN/nm}$ and $\lambda_0=0.017 \mathrm{~pN/nm}$.
This transition is accompanied by the formation of a rigid cylinder, where multiple $z$ values have the same $r$ (\Cref{fig:snapthrough}a).
The abrupt nature of this transition is evident by studying the length as a function of tension for different bending rigidity values (\Cref{fig:snapthrough}b). 
For all values of bending rigidities investigated, we found that decreasing tension resulted in the elongation of the cylinder and the discontinuity indicated the presence of a snapthrough instability.
This discontinuity in length is accompanied by a discontinuity in the energy landscape, such that decreasing tension results in a jump from a high energy state to a low energy state (\Cref{fig:snapthrough}c).
A similar transition can also be seen when the bending rigidity is varied (\Cref{fig:snapthrough}d-f).
The transition from a dome-shaped deformation to a cylinder occurs between $\kappa=77.7 \mathrm{~pN\cdot nm}$ and $\kappa=88.2 \mathrm{~pN\cdot nm}$, while all other parameters are held constant (\Cref{fig:snapthrough}d).
However, the snapthrough transition occurs only for specific values of membrane tension (\Cref{fig:snapthrough}e). 
For example, low tension values result in long tubes smoothly and high tension values result in dome-shaped membranes for a wide range of bending rigidities.
At intermediate values of tension, we observe a snapthrough instability as a discontinuity in the increasing length of the membrane for increasing rigidity and a corresponding drop in energy (\Cref{fig:snapthrough}f).
\begin{figure}[htbp]
    \centering
    \includegraphics[width=\textwidth]{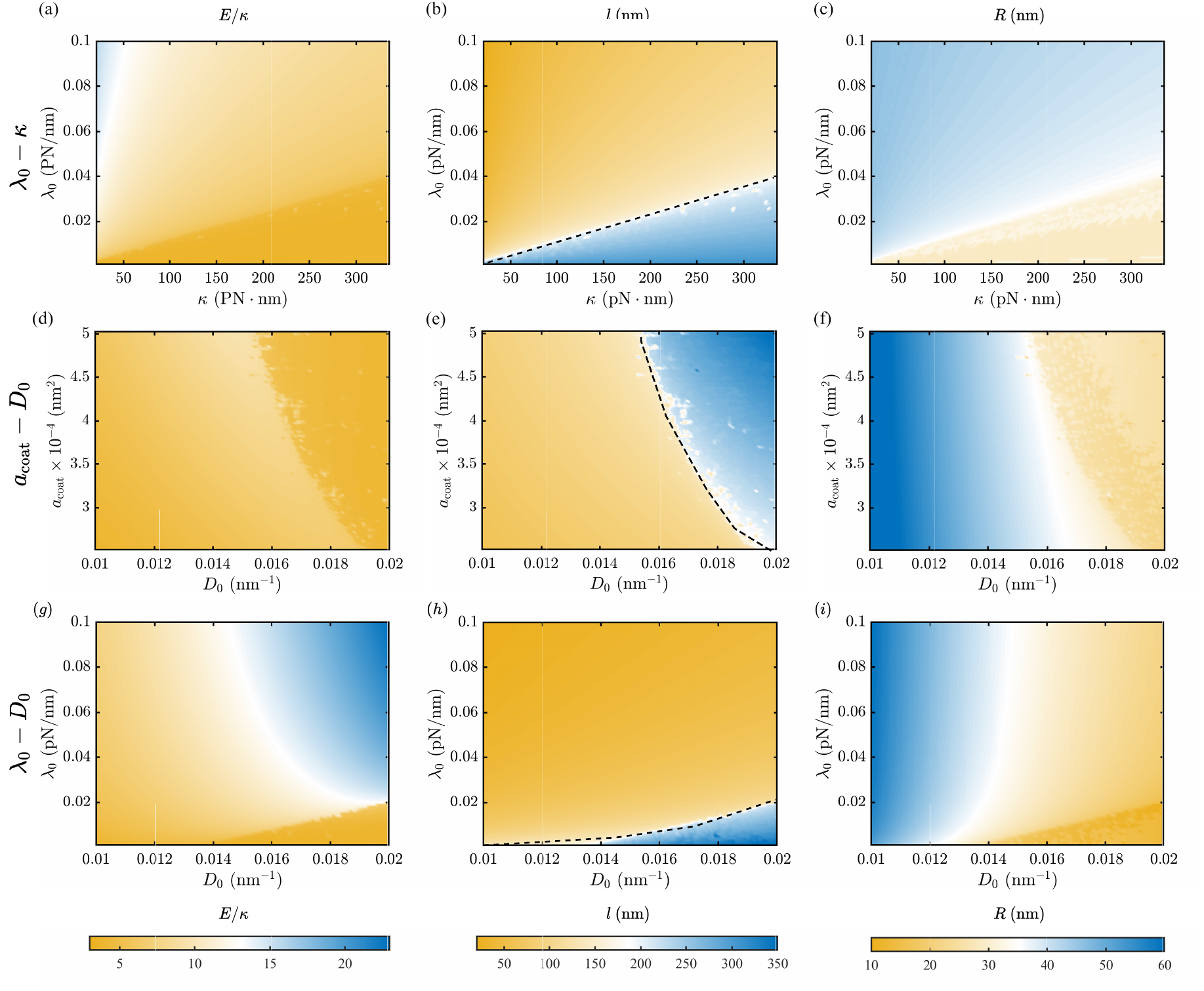}
    \caption{The dome-to-cylinder transition is depicted by the energy landscape in the parameter spaces membrane tension ($\lambda_0$), bending rigidity ($\kappa$), coat area of protein ($a_{\mathrm{coat}}$), and the spontaneous curvatures ($D_0$). We present the phase-space of bending energy, length and radius of the tube in the parameter space of $\lambda_0-\kappa$ (a-c) for $a_{\mathrm{coat}}=37700~\mathrm{nm^2}$ and $D_0=0.015~\mathrm{nm^{-1}}$, $a_{\mathrm{coat}}-D_0$ (d-f) for $\kappa=42~\mathrm{pn \cdot nm}$ and $\lambda_0=0.01~\mathrm{pN/nm}$, and $\lambda_0-D_0$ (g-i) for $a_{\mathrm{coat}}=37700~\mathrm{nm^2}$, $\kappa=42~\mathrm{pN \cdot nm}$. The first column (a,d,f) shows the ratio of bending energy to the bending rigidity. Note that bending energy $E$ varies directly with $\kappa$, therefore for $k=336 ~\mathrm{pN\cdot nm}$ $E$ becomes as high as $3000 ~\mathrm{pN\cdot nm}$ in the parameter space of $\lambda_0-\kappa$, whereas, for other two parameter space maximum value of $E$ is around $1000 ~\mathrm{pN\cdot nm}$. To represent the variation in the same color scale we normalized the bending energy with $\kappa$, and noticed sharp energy change across the dome to cylinder transition, which is much more evident in the phase diagram of the length $l$ as shown with a dashed line (b,e,h).
    However, the change in radius across this length transition is not as dramatic as the tube length; still, we notice a transition, as shown by the color contrast from dark yellow to light yellow (c,f,i).
    } \label{fig:phase_diagram}
\end{figure}

The snapthrough behavior of elastic membranes can be demonstrated by saddle node bifurcations \cite{gomez2017critical}. 
However, the nonlinear nature of the present problem precludes stability analysis since closed-form solutions of the governing equation are not possible. 
We sought to use a numerical approach to demonstrate the snapthrough. 
We expect a pair of stable equilibrium solutions in the neighborhood of the transition to demonstrate the existence of saddle node bifurcations.
We conducted simulations by varying tension in the increasing and decreasing direction and using the solution of the previous parameter value as an initial guess for the current parameter value. As we traverse the tension axis, we find that there are two different solutions for the minimum energy shape of the membrane and there is a lag in the tube length and bending energy when we change $\lambda_0$ in forward and backward directions (\Cref{fig:lag}). 
Thus using numerical solutions, we show that the transition from dome-to-cylinder is indeed a snapthrough instability.

We next investigated the effect of varying the protein-induced curvature $D_0$ and coat area $a_{\mathrm{coat}}$ on the snapthrough transition (\Cref{fig:D0_a0}).
For a fixed coat area, we found that the snapthrough transition occurs for increasing coat curvature (\Cref{fig:D0_a0}a).
We see a sharp increase in tube length (\Cref{fig:D0_a0}b) and a corresponding decrease in bending energy drops at the transition (\Cref{fig:D0_a0}c). 
Similar transitions can be seen for increasing the coat area for a fixed value of protein-induced curvature (\Cref{fig:D0_a0}d). 
In this case, we see that length of the tube has a discontinuity with increasing coat area for a specific value $D_0=0.017 \mathrm{~nm}^{-1}$ (\Cref{fig:D0_a0}e) and this corresponds to a sharp decrease in the bending energy (\Cref{fig:D0_a0}f). 
For the other values of $D_0$ tested, we did not observe any transition --  a cylinder remains as a cylinder and a dome remains a dome in these cases. The corresponding energies are also monotonic.


\subsection{Phase diagrams show transition boundaries for different tube parameters}
Our results thus far indicate that the transition from a dome-shaped deformation to a rigid cylindrical tube is governed by a snapthrough instability that depends on multiple parameters such as protein-induced curvature, bending rigidity, edge tension, and coat area. 
In order to understand how these parameters affect tube length, tube radius, and bending energy, we systematically varied them in physically relevant ranges (\Cref{table:parameters}). 
For each parameter combination, we plot the tube length, tube radius, and the bending energy (\Cref{fig:phase_diagram}).
In the tension-bending rigidity space, the energy landscape shows a curved surface where high bending rigidity has high bending energy but increasing tension lowers the bending energy for all values of bending rigidity (\Cref{fig:phase_diagram}a).
The length of the tube had a smooth and linear transition from short tubes to long tubes (\Cref{fig:phase_diagram}b). 
A similar transition is seen for the tube radius; thicker tubes are formed for higher tension and thinner tubes for higher bending rigidity (\Cref{fig:phase_diagram}c).

In the coat area-protein-induced curvature space, the bending energy landscape is non-monotonic and shows a highly curved manifold where the bending energy increases as coat area increases and remains low as coat curvature increases (\Cref{fig:phase_diagram}d). 
For this combination of parameters, the transition from short to long tubes in nonlinear but consistent with our understanding that higher induced curvature and larger area are compatible with longer tubes (\Cref{fig:phase_diagram}e).
The tube radius shows little to no dependence on coat area but depends on the protein-induced curvature with smaller curvatures resulting in larger tube radii (\Cref{fig:phase_diagram}f).
Finally, we investigate the tension-induced-curvature landscape and find that the tube length remains small for most values of tension for all values of induced-curvature. 
The bending energy remains low for most values of tension and induced curvature, increasing for high values of induced curvature (\Cref{fig:phase_diagram}i). 
Only when the tension value is low do we obtain long tubes (\Cref{fig:phase_diagram}h).
The tube radius has a weak dependence on tension for low and high values of induced curvature (\Cref{fig:phase_diagram}i).
At intermediate values, there is a nonlinear dependence on both tension and induced curvature, consistent with the dependence of tube radius on effective tension (see \Cref{eqn:theoretical} and \Cref{fig:dimensionless}a).

Thus, the formation of a cylindrical tubule due to protein-induced curvature depends on membrane properties and is governed by a snapthrough transition.


\section{Discussion}
In the study, we have presented a theoretical formulation of tube formation due to protein-induced anisotropic curvature and used it to simulate cylindrical tubules without any approximations to boundary conditions. 
In our continuum approach, we accounted for the effect of orientation of proteins by giving the specified values of spontaneous curvature.
Using this model, we identified that the energy landscape of protein-induced tubulation is governed by a snapthrough instability and this instability depends on the membrane tension and bending modulus. 
This snapthrough is associated with a transition from a dome-to-tube transition in the length of the tubule.
Furthermore, we observed a downhill of energy across this transition, suggesting how the bending energy landscape favors the transition of the tube shape.

The idea that deviatoric curvature favors tubular protrusion has been proposed in a series of papers \cite{Walani2014,Walani2015,Perutkova2010}, where a cylindrical membrane is coated with proteins that induce deviatoric curvature. 
However, formation of tube from a flat membrane due to anisotropic curvature alone (in the absence of any localized force at the pole), has been investigated in only a few studies. 
Of note is the study by Perutkova \textit{et al.} \cite{Perutkova2010}, which shows that tubular shapes in the presence of deviatoric curvature are energetically favorable.
In studies focused on modeling yeast endocytosis, where the endocytic invagination forms a tube, localized forces were used to model the tubulation and such forces were ascribed to actin \cite{goode2015actin,Liu2009-sr,kaksonen2005modular}. 
Here we show that anisotropic curvatures such as those induced by BAR-domain proteins are sufficient to form a cylindrical tube. 

While surveying the parameter space for tubulation, we observed that the elongation of a tube was governed by a snapthrough instability.
We previously reported a snapthrough instability for the formation of a spherical bud \cite{Hassinger17} and find that tubulation is also governed by a similar tension-dependent transition. 
Additionally, the equivalent effective tension helps us understand the relationship between tube radius and contributions from the different terms. These relationships have implications in biophysical processes where changes of lipid composition and lipid saturation leads to changes in the effective tension \cite{sitarska2020pay}. 

The predictions from our model have implications understanding the formation and maintenance of tubular structures generated by proteins in cells such as formation of t-tubule in myocytes, cristae formation in mitochondria, and trafficking from intracellular organelles to the plasma membrane. 
The maintenance of t-tubules in muscle cells depends on the distribution of BAR-domain containing proteins called BIN1 among others \cite{vitale2021t}. 
In heart failure, there is a marked reduction of BIN1 and the t-tubule network is significantly modified leading to loss of excitation-contraction coupling \cite{zhou2017cardiac}. 
Similarly, the maintenance of the tubular invagination in yeast during endocytosis depends on the F-BAR and N-BAR proteins and the absences of these proteins disrupts the progression of endocytosis \cite{kishimoto2011determinants}.

We note that our present work has a number of simplifications particularly with respect to the protein coat. 
The binding and unbinding of these anisotropic curvature inducing proteins is curvature dependent \cite{sutton2022artificial}.
The accumulation of these protein dimers and the interaction between their curvature sensing and curvature generation abilities also depends on their adsorption, diffusion, and aggregation in the plane of the membrane \cite{capraro2010curvature}.
Such processes have been studied using molecular dynamics simulations  \cite{arkhipov2009membrane,yin2009simulations} and a continuum description of these process that couples the underlying thermodynamics \cite{das11,frost2008structural} coupled with membrane deformation would be necessary to avoid the model simplifications. 
We also anticipate that the interactions between protein-mediated tubulation and force-mediated tubulation would be necessary to capture cellular processes where both membrane-bound proteins and actin-mediated forces are involved \cite{akamatsu2020principles}.
Finally, relaxation of the assumption of axisymmetry in tubulation could be necesary to investigate how forces and proteins may interact \cite{zhu2022mem3dg,vasan2020mechanical,auddya2021biomembranes}.
In summary, we expect that the findings from our work will motivate the development of these future efforts and gain a deeper understanding of how membrane tubes form. 
    
\section*{Acknowledgments}
The authors would like to thank Profs. David Saintillan and Itay Budin for their critical feedback.
They would also like to thank Dr. Miriam Bell, Dr. Aravind Chandrasekaran, and Mr. Cuncheng Zhu for their input and discussions.
This work was supported by NIH R01GM132106 to P.R.

\clearpage

\appendix

{\LARGE\bfseries Supporting Information}
\renewcommand{\thefigure}{S\arabic{figure}}
\setcounter{figure}{0}

\section{Model Development}
\label{sec:si_sec1}
\subsection{Surface representation}
\label{sec:si_sec1_1}
In a polar coordinate the membrane can be parameterized by the arclength $s$ and the rotation angle $\theta$ as
\begin{myequation}
    \boldsymbol{r}(r,z,\theta)=\boldsymbol{r}(s,\theta).
\end{myequation}
The surface tangents are given by $\boldsymbol{e}_s=\boldsymbol{r}_{,s}$ and $\boldsymbol{r}_{,\theta}$.
The surface metric $a_{ij}=\boldsymbol{e}_i \cdot \boldsymbol{e}_{j}$ becomes
\begin{myequation}
a_{i j}=\left[\begin{array}{ll}
1 & 0 \\
0 & r^{2}
\end{array}\right].
\end{myequation}
The curvature tensor $b_{ij}=e_{i,j}\cdot \boldsymbol{n}$ simplifies to 
\begin{myequation}
b_{i j}=\left[\begin{array}{cc}
\psi_s & 0 \\
0 & r \sin \psi
\end{array}\right].
\end{myequation}
The mean curvatures is given by
$$H=\frac{1}{2}a^{\alpha \beta} b_{\alpha \beta},$$
where, $a^{\alpha \beta}$ is the inverse of the metric tension.
The principal curvature can be extracted from the curvature tensor as
$$c_{\zeta}=b_{\alpha \beta}\zeta^{\alpha}\zeta^{\beta},$$
and $$c_{\mu}=b_{\alpha \beta}\mu^{\alpha}\mu^{\beta}.$$
where, $\boldsymbol{\zeta}$ and $\boldsymbol{\mu}$ are surface tangents in two principal directions,
The deviatoric curvature becomes
$$D=\frac{1}{2}(c_{\zeta}-c_{\mu}).$$



\subsection{Protein Orientation}
\label{sec:si_sec1_2}
\begin{figure}[!h]
    \centering
    \includegraphics[scale=.75]{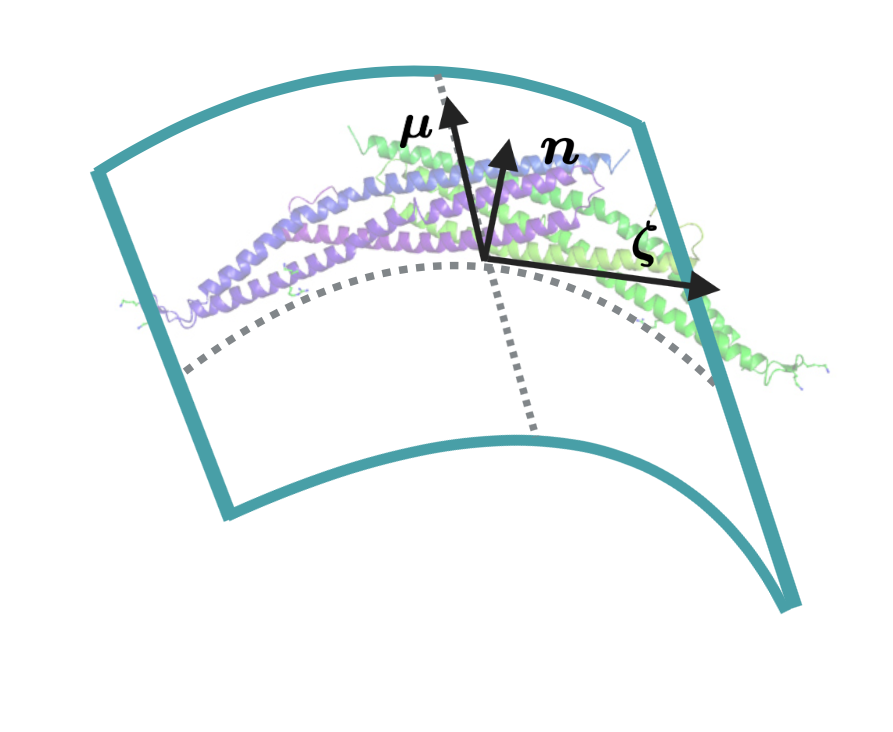}
    \caption{Orientation vectors of BAR-domain proteins}
    \label{fig:Prb1}
\end{figure}
The orientation of a protein on the surface can be represented by orientation unit vector $\boldsymbol{\zeta}$ (\Cref{fig:Prb1}) which essentially indicates tangent to the curve on which protein orients \cite{Walani2014}. Thus we can constitute another unit vector $\boldsymbol{\mu}$, such that: $\boldsymbol{\mu}=\boldsymbol{n}\times\boldsymbol{\zeta}$. 

\subsection{Balance relations}

The force balance equation is dictated by 
\begin{myequation}
 \boldsymbol{T}^{\alpha}_{;\alpha}+p\boldsymbol{n}=\boldsymbol{0}\,,
\label{eqn:eqm}
\end{myequation}%
where $p$ is normal pressure on the membrane and $\boldsymbol{T}$ is traction on the membrane and given by,
\begin{myequation}
\boldsymbol{T}^{\alpha }=N^{\beta \alpha }\boldsymbol{a}_{\beta }+S^{\alpha }\boldsymbol{n}.
\label{eqn:stress_balance_2}
\end{myequation}%
Here, $\boldsymbol{N}$ is the in-plane components of the stress and is given by
\begin{myequation}
 N^{\beta \alpha }=\zeta^{\beta \alpha }+b_{\mu }^{\beta }M^{\mu
\alpha} \qquad \mathrm{and} \qquad S^{\alpha }=-M_{;\beta }^{\alpha \beta },
\label{eqn:stresses}
\end{myequation}
where $\sigma^{\beta \alpha}$ and $M^{\beta \alpha}$ are obtained from the following constitutive relations \cite{Steigmann99}
\begin{myequation}
   \sigma^{\beta \alpha }=\rho \left(\frac{\partial F}{\partial a_{\alpha \beta}}+\frac{\partial F}{\partial a_{\beta \alpha}}\right) \quad \text{and} \quad \boldsymbol{M}^{\beta \alpha }=\frac{\rho}{2} \left(\frac{\partial F}{\partial b_{\alpha \beta}}+\frac{\partial F}{\partial b_{\beta \alpha}}\right),
\end{myequation}
with $F=W/\rho$ as the energy mass density of the membrane.
Combining these we get the balance equations in tangent and normal direction 
\begin{myequation}
\label{eqn:eqns_of_motion_a}
N_{;\alpha }^{\beta \alpha }-S^{\alpha }b_{\alpha }^{\beta }=0, \quad S_{;\alpha }^{\alpha }+N^{\beta \alpha }b_{\beta \alpha }+p=0     
\end{myequation}
The normal force balance relation in Equation \ref{eqn:eqns_of_motion_a}$_{ii}$ becomes \cite{Walani2014}
\begin{myequation}
\label{eqn:walani_shape}
\begin{split}
&\underbrace{\frac{1}{2} [W_D(\zeta^{\alpha}\zeta^{\beta}-\mu^{\alpha}\mu^{\beta})]_{;\beta \alpha}}_{\text{I}} + \underbrace{\frac{1}{2} W_D (\zeta^{\alpha}\zeta^{\beta}-\mu^{\alpha}\mu^{\beta})b_{\alpha \gamma} b^{\gamma}_{\beta}}_{\text{II}}+ \\
&\Delta \left(\frac{1}{2}W_H\right)+ (W_K)_{;\beta \alpha}\left(2H {a}^{\beta \alpha}- {b}^{\beta \alpha} \right)+ W_H(2H^2-K)+2H(KW_K-W)-2H\lambda=p,
\end{split}
\end{myequation}
where the marked terms are simplified in the next section for an axisymmetric geometry.
To construct a force boundary condition we use the expression of the normal traction force as given by \cite{agrawal2009boundary}
\begin{myequation}
\begin{split}
\label{eqn:traction}
F_{n}=&\left(\tau W_{K}\right)^{\prime}-\frac{1}{2}\left(W_{H}\right)_{, \nu}-\left(W_{K}\right)_{, \beta} \tilde{b}^{\alpha \beta} v_{\alpha} \\
&+\frac{1}{2}\left(W_{D}\right)_{, \nu}-\left(W_{D} \lambda^{\alpha} \lambda^{\beta}\right)_{; \beta} v_{\alpha}-\left(W_{D} \lambda^{\alpha} \lambda^{\beta} v_{\beta} \tau_{\alpha}\right)^{\prime},
\end{split}
\end{myequation}
where $\boldsymbol{\tau}$ is the unit tangent to the curve at boundary, $\boldsymbol{\nu}$ is the outward normal to the same curve at the boundary, and can be constructed from local surface normal $\boldsymbol{n}$ as $\boldsymbol{\nu}=\boldsymbol{\tau}\times \boldsymbol{n}$.

\section{Simplification in axisymmetry}
\label{sec:si_sec2}
\subsection{Governing equations}
\label{sec:si_gov_eqn}
We have orthogonal surface tangent vectors as given by
\begin{myequation}
    \boldsymbol{a}_1= \boldsymbol{e}_s, \quad  \boldsymbol{a}_2= r\boldsymbol{e}_{\theta}.
\end{myequation}
We get the expression of orientation unit vector in terms of orthogonal basis vectors as given below
\begin{myequation}
    \boldsymbol{\zeta}=-\boldsymbol{e}_{\theta}=-\frac{1}{r} \boldsymbol{a}_2, \quad  \boldsymbol{\mu}= \boldsymbol{a}_1. 
\end{myequation}
We first find the expressions of the direct products of orientation vectors used in \Cref{eqn:walani_shape} below
\begin{myequation}
\zeta^ {\alpha} \zeta^{\beta}=
\begin{pmatrix}
0 & 0\\
0 & 1/r^2
\end{pmatrix},
\end{myequation}
and
\begin{myequation}
\mu^ {\alpha} \mu^{\beta}=
\begin{pmatrix}
1 & 0\\
0 & 0
\end{pmatrix}.
\end{myequation}

In the limit of axisymmetry, the components of Christoffel symbols denoted by $$\Gamma_{b c}^{a}=\frac{1}{2} a^{a d}\left[\partial_a a_{b d}+\partial_{b} a_{d c} -\partial_d a_{b c}\right].$$ The components of the Christoffel are given below with $1$ and $2$ denoting the arclength ($s$) and azimuthal direction, respectively
\begin{myequation}
\begin{array}{l}
\Gamma_{11}^{1}=0, \quad \Gamma_{22}^{2}=0, \quad \Gamma_{22}^{1}=-r \cos \psi, \\
\Gamma_{12}^{2}=\Gamma_{21}^{2}=\frac{\cos \psi}{r}, \quad \Gamma^1_{21}=0, \text{and} \quad \Gamma_{11}^{2}=0.
\end{array}
\end{myequation}
We first simplify the term $\text{I}$ in \Cref{eqn:walani_shape} below 
\begin{myequation}
\label{eqn:termIs}
\begin{split}
        \text{I} &= \frac{1}{2}\left[W_{D}\left(\zeta^{\alpha} \zeta^{\beta}-\mu^{\alpha} \mu^{\beta}\right)\right]_{; \beta \alpha} \\
        &=(W_D \zeta^{\alpha} \zeta^{\beta})_{;\beta \alpha} -\frac{1}{2}\left[W_{D}\left(\zeta^{\alpha} \zeta^{\beta}+\mu^{\alpha} \mu^{\beta}\right)\right]_{; \beta \alpha} \\
        & = (W_D \zeta^{\alpha} \zeta^{\beta})_{;\beta \alpha} -\frac{1}{2} (W_D a^{\alpha \beta })_{; \beta \alpha }. 
\end{split}
\end{myequation}
Note that we recover the surface metric from the addition of the direct products of the orientation vectors as given below 
\begin{myequation}
\left(\zeta^{\alpha} \zeta^{\beta}+\mu^{\alpha} \mu^{\beta}\right)=\left(\begin{array}{ll}
1 & 0 \\
0 & 1 / r^{2}
\end{array}\right) = a^{\alpha \beta}.
\end{myequation}
From Equation \ref{eqn:termIs} we can further write term I as
\begin{myequation}
\begin{split}
     \text{I}& = (W_D \zeta^{\alpha} \zeta^{\beta})_{;\beta \alpha} -\frac{1}{2} \Delta (W_D) \\
     & = \eta^{\beta}_{;\beta} -\frac{1}{2} \Delta (W_D),
\end{split}
\end{myequation}
where
\begin{myequation}
\label{eqn:termI_eta}
\begin{aligned}
&\eta^{\beta}=\left(W_{D} \zeta^{\alpha} \zeta^{\beta}\right)_ {; \alpha}\\
&=\left(W_D \zeta^{\alpha} \zeta^{\beta} \right)_{,\alpha} +W_D\Gamma_{\alpha \gamma}^{\alpha} \zeta^{\gamma} \zeta^{\beta}+W_D \Gamma_{\alpha \gamma}^{\beta} \zeta^{\alpha}  \zeta^{\gamma}.
\end{aligned}
\end{myequation}
The components of $\eta^\beta$ are estimated below in two principal directions 
\begin{myequation}
    \eta^1=0+0+ W_D\Gamma ^1_{2 2} \zeta^2 \zeta^2 =-\frac{\cos \psi}{r} W_D,
\end{myequation}
and
\begin{myequation}
    \eta^2=0+0+0=0.
\end{myequation}
The divergent $\eta^{\beta}_{;\beta}$ reduces to 
\begin{myequation}
\begin{split}
    \eta^{\beta}_{;\beta} &= \frac{1}{\sqrt{a}} (\sqrt{a} \eta^\beta )_{,\beta} \\
    & =  \frac{1}{r} (r \eta^1 )_{,1} \\
    & = - \frac{(\cos \psi W_D)^{\prime}}{r}.
\end{split}
\end{myequation}
Substituting the expression of $\eta^{\beta}_{;\beta}$ in Equation \ref{eqn:termIs} we get term I simplified as 
\begin{myequation}
\label{eqn:termI_simp}
    \text{I}=-\frac{1}{2} \Delta (W_D)-\frac{(\cos \psi W_D )^{\prime}}{r}.
\end{myequation}

Next, we simplify term II below 
\begin{myequation}
\begin{split}
\label{eqn:termII_simp}
    \text{II}&=\frac{1}{2} W_{D}\left(\zeta^{\alpha} \zeta^{\beta}-\mu^{\alpha} \mu^{\beta}\right) b_{\alpha \gamma} b_{\beta}^{\gamma} \\
    &= \frac{1}{2} W_{D} \left\{ \zeta^2 \zeta^2 b_{22}b^2_2-\mu^1 \mu^1 b_{11}b^1_1 \right\} \\
    &= \frac{1}{2}W_D \left\{ \frac{\sin^2 \psi}{r^2} -{\psi^{\prime}}^2 \right\} \\
    &=  \frac{1}{2}W_D \left( \frac{\sin \psi}{r} +{\psi^{\prime}} \right) \left( \frac{\sin \psi}{r} -{\psi^{\prime}} \right) \\
    &= \frac{1}{2}W_D ~ 2H ~ 2D \\
    & =2HDW_D.
\end{split}
\end{myequation}
Finally, using the simplifications of term I (\Cref{eqn:termI_simp}) and term II (\Cref{eqn:termII_simp}), the shape equation becomes
\begin{myequation}
\begin{aligned}
p=& \frac{{L}^{\prime}}{r}+W_{H}\left(2 H^{2}-K\right)-2 H\left(W+\lambda-W_{D} D\right),
\end{aligned}
\end{myequation}
where $L$ relates to the expression of the traction as shown in \Cref{eqn:traction}, given by
\begin{myequation}
\label{eqn:L_def}
{L} / r=\frac{1}{2}\left[\left(W_{H}\right)^{\prime}-\left(W_{D}\right)^{\prime}\right] -\frac{\cos \psi}{r} W_D= -F_n. 
\end{myequation}
The above relation gives a natural boundary condition for $L$ at the center and the boundary. At the center it directly correlates with the value of pulling force as
\begin{myequation}
    p_f= \lim_{r \xrightarrow{} 0} 2 \pi r F_n = - 2 \pi {L}(0).
\end{myequation}
Note than the derivation of shape equation was presented in \cite{Walani2014} where the last term was missing in the definition of $L/r$ in \Cref{eqn:L_def} and which led to an incorrect residual term $\frac{\left(W_{D}\right)^{\prime} \cos \psi}{r}$ in the shape equation. Please note that an artificial pulling force was introduced for the boundary condition of $L=0$ at the center of the membrane with the incomplete expression as presented in Walani et al. \cite{Walani2014}.

\subsection{Area parameterization}
The governing equation is solved on a patch of membrane with fixed surface area, where the coat area of protein is prescribed.
The arclength parametrization poses some difficulty since total arclength varies depending on the equilibrium shape of the membrane. 
Therefore, we did a coordinate transformation of arclength to a local area $a$ as given by
\begin{myequation}
    \frac{\partial }{\partial s} =2 \pi r\frac{\partial }{\partial a}. 
\end{myequation}
Note that in the differential form local area relates as
\begin{myequation}
    da=2 \pi {r} d{s}
\end{myequation}
The tangential force balance relation in Equation 7 transforms to
\begin{myequation}
\label{eqn:tang_force_bal_area}
\frac{\partial \lambda}{\partial a}=2 \kappa (H-C_0)\frac{\partial C_0}{\partial a}+2 \kappa_d (D-D_0)\frac{\partial D_0}{\partial a}. 
\end{myequation}
The normal force balance relation in Equation 8 becomes
\begin{myequation}
\begin{aligned}
\label{eqn:normal_force_bal_area}
p=& 2 \pi \frac{\partial L}{\partial a}+2 \kappa (H-C_0)\left(2 H^{2}-K\right)-2 H\left(W+\lambda-2 \kappa_d D(D-D_0)\right)
\end{aligned}
\end{myequation}
where,
\begin{myequation}
\label{eqn:L}
\frac{L}{r}=\pi r \frac{\partial }{\partial a}\bigg\{\kappa (H-C_0)-\kappa_d (D-D_0)\bigg\} - 2 \kappa_D (D-D_0)\frac{\cos \psi}{r}. 
\end{myequation}

\subsection{Non-dimensionalization}
In this section we use $\tilde{(\cdot)}$ to represent the dimensionless quantities.
We used a scale of curvature $1/R_0$, where $R_0$ is the equivalent lengthscale in the domain. The dimensionless mean, deviatoric and Gaussian curvature becomes $\tilde{H}=R_0 H $, $\tilde{D}=R_0 D$, and $\tilde{K}=R_0^2 K$. The same scale for curvature is used to nondimensionalize spontaneous mean and deviatoric curvature and they become $\tilde{C}_0=R_0 C_0$ and $\tilde{D}_0=R_0 D_0$. 
The area is dimensionalized with scale $A_0=2\pi R_0^2$.
The scale for membrane tension is taken as ${\kappa}/{R_0^2}$, therefore $\tilde{\lambda}= {R_0^2 \lambda}/{\kappa}$. The dimensionless form of L becomes $\tilde{L}={R_0 L}/{\kappa}$. 
\\
The tangential force balance relation in \Cref{eqn:tang_force_bal_area} reads as
\begin{myequation}
\label{myeqn:tang_force_bal_axi_nd}
\frac{ \partial \tilde{\lambda}}{\partial \tilde{a}}=2 (\tilde{H}-\tilde{C}_0)\frac{ \partial \tilde{C}_0}{\partial \tilde{a}}   +2 \tilde{\kappa}_d (\tilde{D}-\tilde{D}_0) \frac{ \partial \tilde{D}_0}{\partial \tilde{a}}, 
\end{myequation}
where $\tilde{\kappa}_d=\frac{\kappa_d}{\kappa}$ represents the dimensionless deviatoric curvature. 
The normal force balance relation in \Cref{eqn:normal_force_bal_area} simplifies to
\begin{myequation}
\begin{split}
\tilde{p}= \frac{\partial \tilde{L}}{\partial \tilde{a}}+2 (\tilde{H}-\tilde{C}_0)\left(2 \tilde{H}^{2}-\tilde{K}\right)-&2 \tilde{H}\bigg\{ (\tilde{H}-\tilde{C}_0)^2+\tilde{\kappa}_d(\tilde{D}-\tilde{D}_0)^2 \\
&+\tilde{\lambda}-2 \tilde{\kappa}_d \tilde{D}(\tilde{D}-\tilde{D}_0)\bigg\},
\end{split}
\end{myequation}
with
\begin{myequation}
\label{myeqn:L}
\frac{\tilde{L}}{  \tilde{r}}= \tilde{r}^2 \frac{\partial }{\partial \tilde{a}}\bigg\{ (H-C_0)-\tilde{\kappa}_d (D-D_0)\bigg\} - 2 \tilde{\kappa}_d (D-D_0)\frac{\cos \psi}{r}. 
\end{myequation}




\section{Supplementary Figures}
\begin{figure}[htbp]
    \centering
    \includegraphics[width=\textwidth]{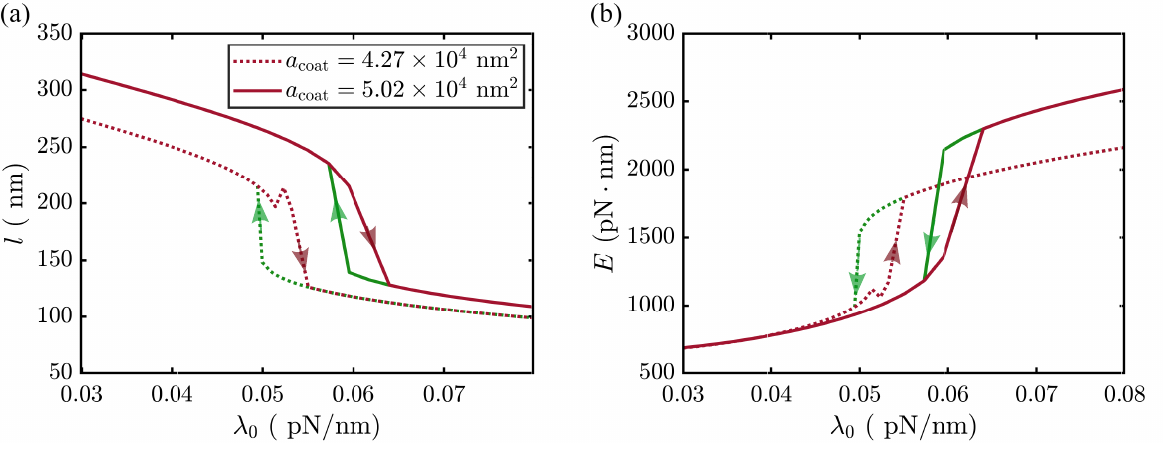}
    \caption{Forward and backward transition to demonstrate snapthrough instability. (a) Transition of tube length in the direction of increasing and decreasing membrane tension for $D_0=0.017~ \mathrm{nm}^{-1}$ and $\kappa=168~ \mathrm{pN\cdot nm}$  and two different values of coat area of proteins. (b) Transition of bending energy in the direction of increasing and decreasing membrane tension for $D_0=0.017~ \mathrm{nm}^{-1}$ and $\kappa=168~ \mathrm{pN\cdot nm}$  and two different values of coat area of proteins.}   \label{fig:lag}
\end{figure}


\printbibliography

\end{document}